\input harvmac

\newif\ifdraft\draftfalse
\newif\ifinter\interfalse
\ifdraft\draftmode\else\interfalse\fi
\def\journal#1&#2(#3){\unskip, \sl #1\ \bf #2 \rm(19#3) }
\def\andjournal#1&#2(#3){\sl #1~\bf #2 \rm (19#3) }

\def\ie{{\it i.e.}}
\def\eg{{\it e.g.}}

\def\p{\partial}

\def\frac#1#2{{#1\over#2}}

\def\inbar{\,\vrule height1.5ex width.4pt depth0pt}
\def\IC{\relax\hbox{$\inbar\kern-.3em{\rm C}$}}
\def\IR{\relax{\rm I\kern-.18em R}}
\def\IP{\relax{\rm I\kern-.18em P}}
\def\Z{{\bf Z}}

%
%


%
\catcode`\@=11
\def\slash#1{\mathord{\mathpalette\c@ncel{#1}}}
\overfullrule=0pt

\def\underrel#1\over#2{\mathrel{\mathop{\kern\z@#1}\limits_{#2}}}

\catcode`\@=12


%

\def \cosh{{\rm cosh}}

\def\exp{{\rm exp}}

\def\ch{{\rm cosh}}


\def\[{[}
\def\]{]}

\def\comment#1{ }

%
\def\draftnote#1{\ifdraft{\baselineskip2ex
                 \vbox{\kern1em\hrule\hbox{\vrule\kern1em\vbox{\kern1ex
                 \noindent \underbar{NOTE}: #1
             \vskip1ex}\kern1em\vrule}\hrule}}\fi}
\def\internote#1{\ifinter{\baselineskip2ex
                 \vbox{\kern1em\hrule\hbox{\vrule\kern1em\vbox{\kern1ex
                 \noindent \underbar{Internal Note}: #1
             \vskip1ex}\kern1em\vrule}\hrule}}\fi}

%
%
\def\al{\alpha}






               

%
%
\def\inbar{\hskip.3em\vrule height1.5ex width.4pt depth0pt}
\def\IC{\relax{\inbar\kern-.3em{\rm C}}}
\def\IN{\relax{\rm I\kern-.16em N}}
\def\IQ{\relax\hbox{$\inbar$\kern-.3em{\rm Q}}}
\def\IZ{\relax{\rm Z\kern-.8em Z}}
%
%

%

\def\g{g_s}
\def\al{{\alpha^\prime}}
\def\bh{\beta_H}
\def\b{\beta}
%
\rightline{EFI-2000-54}
\Title{
\rightline{hep-th/0012258}}
{\vbox{\centerline{Comments on the Thermodynamics}
\vskip 10pt
\centerline{of Little String Theory}}}
\bigskip
\centerline{D. Kutasov and D. A. Sahakyan}
\bigskip
\centerline{\it Department of Physics, University of Chicago}
\centerline{\it 5640 S. Ellis Av., Chicago, IL 60637, USA}
\centerline{kutasov, sahakian@theory.uchicago.edu}

\bigskip
\noindent
We study the high energy thermodynamics of Little String Theory,
using its holographic description. This leads to the entropy-energy 
relation $S=\beta_H E+\alpha\log E+O(1/E)$. We compute $\alpha$ and 
show that it is negative; as a consequence, the high energy 
thermodynamics is unstable. We exhibit a mode localized near the 
horizon of the black brane, which has winding number one around 
Euclidean time and a mass that vanishes at large  $E$ (or 
$\beta\to\beta_H$). We argue that the high temperature phase of 
the theory involves condensation of this mode.

\vfill

\Date{December 26, 2000}

\lref\berk{M. Berkooz and M. Rozali, ``Near Hagedorn Dynamics of NS
Fivebranes or A New Universality Class of Coiled Strings'', hep-th/0005047.}

\lref\har{T. Harmark and N. A. Obers, ``Hagedorn Behaviour of Little String 
Theory from String Corrections to NS5-Branes'', hep-th/0005021.}

\lref\Seib{N. Seiberg, ``New theories in six dimensions and matrix description 
of M-theory on $T^5$ and $T^5/Z_2$'', Phys. Lett. {\bf B408} (1997) 98,
hep-th/9705221.}

\lref\abks{O. Aharony, M. Berkooz, D. Kutasov and N. Seiberg, 
``Linear Dilatons, NS5-branes and Holography'', 
JHEP {\bf 9810} (1998) 004, hep-th/9808149.}

\lref\mald{J. M. Maldacena, ``Statistical entropy of near-extremal fivebranes'',
Nucl. Phys. {\bf B477} (1996) 168, hep-th/9605016.}

\lref\polch{J. Polchinski, ``Evaluation of the One Loop String Path Integral'',
Commun. Math. Phys. {\bf 104} (1986) 37.}

\lref\att{J. J. Atick and E. Witten, ``The Hagedorn Transition and the Number of Degrees
of Freedom of String Theory'', Nucl. Phys. {\bf B310} (1988) 291.}

\lref\difra{P. Di Franchesko, P. Mathieu, D. Senechal, ``Conformal Field
Theory'', Springer-Verlag, 1997}

\lref\polchin{J. Polchinski, ``String theory'', vol. 1, Cambridge University
Press, 1998.}

\lref\aharon{O. Aharony, ``A brief review of little string theories'',
Class. Quant. Grav. {\bf 17} (2000) 929, hep-th/9911147.}

\lref\mc{B. McClain and B. D. B. Roth, ``Modular Invariance for Interacting
Bosonic String at Finite Temperature'', Commun. Math. Phys. {\bf 111} (1987)
539.}

\lref\brien{K. H. O'Brien and C.-I. Tan, ``Modular Invariance of the
thermo-partition function and global phase structure of the heterotic 
string'', Phys. Rev. {\bf D36} (1987) 1184.}  

\lref\GKP{A. Giveon, D. Kutasov and O. Pelc, ``Holography for Non-Critical
Superstrings'', JHEP {\bf 9910} (1999) 035, hep-th/9907178.}

\lref\GK{A. Giveon and D. Kutasov, ``Little String Theory in a Double
Scaling Limit'', JHEP {\bf 9910} (1999) 034, hep-th/9909110; 
``Comments on Double Scaled Little String Theory'', JHEP {\bf 0001} (2000)
023, hep-th/9911039.}

\lref\HS{G. Horowitz and A. Strominger, ``Black Strings and $p$--Branes'',
Nucl. Phys. {\bf B360} (1991) 197; 
J. Maldacena and A. Strominger, ``Semiclassical decay of near extremal
fivebranes'', JHEP {\bf 9712} (1997) 008, hep-th/9710014.}

\lref\TESCH{J. Teschner, ``On structure constants and fusion rules in the
$SL(2, C)/SU(2)$ WZNW model'',  Nucl. Phys. {\bf B546} (1999) 390,
hep-th/9712256;  
``The Mini-Superspace Limit of the $SL(2,C)/SU(2)$-WZNW Model'', Nucl. Phys.
{\bf B546} (1999) 369, hep-th/9712258; 
``Operator product expansion and factorization in the $H_3^+$-WZNW model'',
Nucl. Phys. {\bf B571} (2000) 555, hep-th/9906215.}

\lref\GKS{A. Giveon, D. Kutasov and N. Seiberg, 
``Comments on String Theory on $AdS_3$'', Adv. Theor. Math. Phys. {\bf 2}
(1998) 733, hep-th/9806194.}

\lref\ks{D. Kutasov and N. Seiberg, ``Noncritical Superstrings'', Phys. Lett.
{\bf 251B} (1990) 67; 
D. Kutasov, ``Some Properties of (Non) Critical Strings'', hep-th/9110041.}

\lref\kazsuz{Y. Kazama and H. Suzuki, 
``New $N=2$ Superconformal Field Theories and Superstring Compactifications'',
Nucl. Phys. {\bf B321} (1989) 232.}

\lref\gpy{D. J. Gross, M. J. Perry and L. G. Yaffe, 
``Instability of Flat Space at Finite Temperature'', 
Phys. Rev. {\bf D25} (1982) 330.}

\newsec{Introduction}

Little String Theory (LST) \aharon\ is a 
non-local theory without gravity which appears in two related contexts. 
One involves the study of vacua of string theory which contain Neveu-Schwarz
(NS) fivebranes in the decoupling limit $g_s\to 0$. In this limit all string 
modes that live in the bulk of spacetime decouple, but the physics on
the fivebranes remains non-trivial \Seib. This gives rise to an interacting
theory in $5+1$ or less dimensions with sixteen or fewer supercharges.

An alternative definition of LST involves string theory on
Calabi-Yau (CY) spaces, at points in the moduli space of vacua where the
CY space develops an isolated singularity \GKP. Sending $g_s\to 0$
one again finds trivial physics in the bulk of the CY manifold, with
interacting physics at the singular point.

In this paper we focus on the maximally supersymmetric
vacuum of LST in $5+1$ dimensions, corresponding to $N$ flat parallel 
$NS5$-branes with worldvolume $\IR^{5,1}$. Most of the analysis
generalizes trivially to the large class of vacua of LST constructed
in \GKP, but we will not discuss the details here.

Holography relates LST to string theory in the near-horizon geometry
of the fivebranes \abks. An important source of difficulty in studying
detailed properties of the theory using this approach is the fact that the
near-horizon geometry includes a ``linear dilaton'' direction, the real line
$\IR_\phi$ (which we will label by $\phi$) along which the dilaton $\Phi$
varies linearly, 
\eqn\lindil{\Phi=-{Q\over2}\phi,}
with $Q$ a model-dependent constant. The behavior \lindil\ implies that
the string coupling $\exp(\Phi)=\exp(-Q\phi/2)$ vanishes as $\phi\to\infty$
(the region far from the fivebranes), but it diverges as one approaches the
fivebranes. Thus, the study of LST using holography involves solving the
dual theory at strong coupling. 

There are some situations in which the strong coupling problem
mentioned above can be avoided. For example, a large class of observables in
the theory can be identified by analyzing non-normalizable vertex operators
whose wave-functions are exponentially supported in the weak coupling 
region $\phi\to\infty$ \abks. Also, one can study LST along its
moduli space of vacua \GK, since in such situations the strong coupling
singularity  associated with \lindil\ is eliminated and one can study the
theory in  a weak coupling expansion. 

In this paper we discuss another situation where LST becomes
weakly coupled and thus amenable to a perturbative holographic 
analysis -- the high energy density regime. As we review in 
section 2, LST has a Hagedorn density of states at very high energies, 
\eqn\hagspec{\rho(E)=e^{S(E)}\sim e^{\beta_H E}.}
The entropy is linear in the energy, and the
inverse temperature $\beta$ is constant,
\eqn\invtemp{\beta={\partial S\over\partial E}=\beta_H.}
The thermodynamics is thus degenerate for very large energy,
and it is of interest to compute finite energy corrections 
to \hagspec, \invtemp. We do this in section 3 and find that, as 
suggested in \refs{\berk,\har}, the density of states has the form 
\eqn\enentr{\rho(E)\sim E^\alpha e^{\beta_H E}
\left[1+O\left({1\over E}\right)\right].}
One of our main purposes is to compute the constant $\alpha$.
We find that $\alpha$ is negative and therefore the high energy
thermodynamics is unstable.  The density of states \enentr\ 
gives rise to the temperature-energy relation
\eqn\tempener{
\beta={\p\log\rho\over\p E}=\beta_H+{\alpha\over E}+
O\left({1\over E^2}\right).}
Since $\alpha$ is negative, the temperature is above the Hagedorn
temperature $T_H=1/\beta_H$, and the specific heat is negative --
increasing the energy of the system leads to a decrease
of the temperature. This behavior is reminiscent of black holes
in flat spacetime, which also have negative specific heat. In 
section 4 we argue that LST in fact undergoes a phase transition
at or around the Hagedorn temperature $T_H$. In a Euclidean
time representation of the thermodynamics, a mode with winding
number one around Euclidean time goes to zero mass at $T_H$. 
It is likely that it becomes tachyonic above the Hagedorn 
temperature and condenses. 

\newsec{Classical NS fivebrane thermodynamics}

The supergravity solution for $N$ coincident near-extremal $NS5$-branes 
in the string frame is \HS:
\eqn\metr{
ds^2=-\left(1-{r_0^2\over r^2}\right)dt^2+\left(1+{N\al\over r^2}\right)
\left({dr^2\over 1-{r_0^2\over r^2}}+r^2d\Omega_3^2\right)+dy_5^2,}
\eqn\dil{
e^{2\Phi}=\g^2(1+{N\al\over r^2}).}
$r=r_0$ is the location of the horizon, $dy_5^2$ denotes the flat
metric along the fivebranes, and $d\Omega_3^2$ is the metric on a
unit three-sphere. The solution also 
involves a non-zero $NS$ $B_{\mu\nu}$ field which we suppress.  
The configuration \metr, \dil\ has energy per unit volume 
\eqn\ener{{E\over V_5}={1\over (2\pi)^5\al^3}\left({N\over\g^2}+\mu\right),}
where 
\eqn\mmm{\mu={r_0^2\over\g^2\al}.}
The first term in \ener\ is the tension of extremal $NS5$-branes 
and can be ignored for the thermodynamic considerations below (it 
is a ground state energy). $\mu$ measures the energy density above
extremality and $\g$ is the asymptotic string coupling, which goes 
to zero in the decoupling limit. 

\lref\witten{E. Witten, ``On String Theory and Black Holes'',
Phys. Rev. {\bf D44} (1991) 314.}

The near-horizon geometry is obtained by sending $r_0, \g\to 0$ keeping
the energy density $\mu$ fixed. Changing coordinates to $r=r_0 
\cosh\sigma$ and Wick rotating $t\to it$ to study the 
thermodynamics, one finds 
\eqn\metro{ds^2=\tanh^2\sigma dt^2+N\al d\sigma^2+N\al d\Omega_3^2+dy_5^2,}
\eqn\dila{e^{2\Phi}={N\over\mu\cosh^2\sigma}.}
String propagation in this geometry corresponds to an ``exact conformal
field theory'',
\eqn\thermback{H_3^+/ U(1)\times SU(2)_N\times \IR^5,}
where 
\eqn\hthreepl{H_3^+={SL(2,C)_N\over SU(2)_N}}
is the Euclidean $AdS_3$ CFT which plays an important role
in the AdS-CFT correspondence (see \eg\ \refs{\TESCH, \GKS}); 
the coset $H_3^+/U(1)$, parametrized by $(\sigma, t)$ in
\metro, is a semi-infinite cigar \witten. The second factor in \thermback\
describes the angular three-sphere.\foot{As is well known, CFT on a 
three-sphere 
with a suitable NS $B_{\mu\nu}$ field is described by the $SU(2)$ 
WZW model.} The radius of the three-sphere can be read off \metro,
\eqn\rsphere{R_{\rm sphere}=\sqrt{N\alpha'}.}
Finally, the third factor in \thermback\ describes the spatial
directions along the fivebranes.

The number of fivebranes $N$ determines the levels of the $SL(2)$
and $SU(2)$ current algebras in \thermback, \hthreepl. More precisely,
since \thermback\ is a background for the superstring, the worldsheet
theory contains fermions; the total level $N$ of the $SU(2)$ and $SL(2)$
current algebras receives a contribution of $N-2$ and $N+2$ respectively
from the bosons, and $+2$ and $-2$ respectively from the fermions.
The total central charge is
\eqn\totcentr{{3(N-2)\over N}+{3(N+2)\over
N}+5+8\cdot{1\over2}=15,} 
which is the correct value for the superstring. Note
also that  the background \thermback\ is exact so we do not have to worry about
$\alpha'$ (or, equivalently, $1/N$) corrections to the geometry. 

The absence of a conical singularity at the tip  
($\sigma=0$ in \metro) requires the circumference  of the cigar to be
\eqn\temp{\bh=2\pi\sqrt{N\al}.}
Thus, Euclidean time lives on a circle of radius $\sqrt{N\al}$,
and the temperature of the system is $T_H=1/\beta_H$. In particular,
the temperature is independent of the energy density $\mu$, which
determines the value of the string coupling at the tip of the cigar
\dila.

\lref\gpnp{G. Gibbons and M. Perry, ``The Physics of 2-d Stringy
Spacetimes'', Int. J. Mod. Phys. {\bf D1} (1992) 335, hep-th/9204090;
C. R. Nappi and A. Pasquinucci, ``Thermodynamics of Two-Dimensional
Black-Holes'', Mod. Phys. Lett. {\bf A7} (1992) 3337, gr-qc/9208002.}

The fact that the temperature is independent of the energy means
that the entropy is proportional to the energy (see \invtemp). 
Therefore, the free energy is expected to vanish\foot{See \gpnp\
for a related discussion in the low energy gravity approximation.},
\eqn\betaf{-\beta \CF=S-\beta E=0.}
In general in string theory the free energy 
is related to the string partition sum via 
\eqn\cantherm{-\beta\CF\equiv\log Z(\beta)=Z_{\rm string},}
where $Z_{\rm string}$ is the single string partition sum, given
by a sum over connected Riemann surfaces \polch. The string path integral
should be performed over geometries in which Euclidean time is
compactified on a circle of radius $R=\beta/2\pi$ (asymptotically).
For high energies one expects the thermodynamics to be dominated
by the black brane geometry \metr, \metro\ and thus the free energy
is proportional to the partition sum of string theory in the background
\thermback.  

The string partition sum $Z_{\rm string}$ can be expanded as follows:
\eqn\zstrexp{Z_{\rm string}=e^{-2\Phi_0} Z_0+Z_1+e^{2\Phi_0}Z_2+\cdots,}
where $\exp(\Phi_0)$ is the effective string coupling in the geometry
\metro\ and $Z_h$ the genus $h$ partition sum in the background \thermback.
Although the string coupling varies along the cigar (see \dila), it 
is bounded from above by its value at the tip,
\eqn\strnnn{e^{2\Phi_0}={N\over \mu}.}
Therefore, it is natural to associate \strnnn\ with the effective
coupling in \zstrexp. We see that the string coupling expansion
in the background \thermback\ provides an asymptotic expansion
of the free energy in powers of $1/\mu$.

The leading term in the free energy \cantherm, \zstrexp\ goes like 
\eqn\flead{-\beta\CF={\mu\over N}Z_0}
and corresponds to a free energy that goes like the energy ($Z_0$
is proportional to the volume of the fivebrane). 
This term is expected to vanish (see \betaf), and therefore we 
conclude that the spherical partition sum in the background \thermback\
should vanish. The fact that this is indeed the case follows from the 
results of \ks, who analyzed a closely related problem, the 
partition sum of vacua including $N=2$ Liouville as a factor.
$N=2$ Liouville is relevant for our problem since, as argued in \GK,
it is equivalent to the Euclidean cigar SCFT. Also, the reasoning of
\ks\ can be applied directly to the cigar. We will next briefly 
review the argument of \ks\ for the vanishing of $Z_0$, first for
the cigar and then for $N=2$ Liouville.

Normally in string theory the partition sum on the sphere is said 
to vanish, due to the volume of the Conformal Killing Group (CKG)
of the sphere, $SL(2,C)$. If the target space is non-compact, the 
partition sum is actually proportional to $V/{\rm vol}(SL(2,C))$ 
where $V$ is the divergent volume of spacetime. Thus, at first sight 
the partition sum is $\infty/\infty$, \ie\ ill defined. However, in 
most situations one is actually interested in the partition sum per 
unit volume. E.g. if the vacuum is translationally invariant in 
the non-compact directions, the partition sum per unit volume is the 
Lagrangian density in this vacuum (\ie\ the classical cosmological 
constant), and it vanishes due to the volume of the CKG. 

In the Euclidean cigar background $H_3^+/U(1)$, the above discussion has 
to be  reexamined. There is no longer a reason to divide by the
volume of the cigar: the background is not translationally invariant
in $\phi$, and in any case, the holographic prescription of \abks\
relates the free energy to the full string partition sum and not
to the partition sum per unit volume (see \cantherm).

The ratio of the volume of the cigar to the volume of the CKG is
finite in this case. The volume of $H_3^+$ contains precisely the same
kind of divergence as that of the CKG. Since the volumes of the $SU(2)$ 
and $U(1)$ in \thermback, \hthreepl\ are finite, we conclude that the
partition sum of string theory in the cigar background is non-zero.

This conclusion is indeed correct in the bosonic string;
in the superstring one has to take into account the fermionic
zero modes. The CKG of the sphere is generalized to a 
Superconformal Killing Group (SCKG), but
the added zero modes are cancelled by fermionic zero modes of the 
$N=1$ SCFT on the cigar. Thus, it appears that in the superstring as 
well the partition sum in the background \thermback\ is finite, in 
contradiction to \betaf.

This conclusion is incorrect because of an interesting property of
the $N=1$ SCFT on the cigar $H_3^+/U(1)$. It turns out that this
model has an ``accidental'' global $N=2$ superconformal symmetry.
In fact it is a special case of the Kazama-Suzuki construction \kazsuz.  
Thus, the SCFT has twice as many fermionic zero modes as one would 
naively guess, and integrating over them leads to the vanishing of the
spherical partition sum \ks. 

The above argument was made in the language of the SCFT on the cigar.
In the dual \GK\ $N=2$ Liouville theory, the vanishing of the spherical
partition sum can be alternatively exhibited as follows. The $N=2$
Liouville Lagrangian is
\eqn\ntwoliouv{\CL=\CL_0+\lambda\int d^2\theta e^{-{2\over \al Q}(\phi+ix)}+
\bar\lambda\int d^2\bar\theta e^{-{2\over \al Q}(\phi-ix)}.}
$\CL_0$ is the free field Lagrangian for $\phi,x$ in a linear
dilaton background \lindil. The linear dilaton slope $Q$ can be
determined by comparing the central charge of \ntwoliouv, 
$c=3+(3\alpha' Q^2/2)$ to that of the cigar CFT \hthreepl,
$c=3+(6/N)$. This leads to 
\eqn\QQQ{Q=2/\sqrt{N\al}.}
$\phi$ is a rescaled version of $\sigma$; far from the tip of the
cigar one has $\phi\simeq\sqrt{N\al}\sigma$. $x$ can be thought of as the
T-dual of $\theta$; it lives on a circle of radius $\sqrt{\alpha'/N}$. 
$\lambda$ is the $N=2$ Liouville coupling. 
$\int d^2\theta$ is an integral over half of the worldsheet
superspace, $\int d^2\theta=G_{-1/2}^+\bar G_{-1/2}^+$.
The operator $e^{-{2\over \alpha'Q}(\phi+ix)}$ is chiral, \ie\ 
it is annihilated by $G^-$, $\bar G^-$.

A standard scaling analysis shows that the partition sum on the
sphere goes like
\eqn\zzero{Z_0\sim (\lambda\bar\lambda)^{1\over N}.}
Thus, to compute $Z_0$ we can differentiate a number of times w.r.t.
$\lambda$, $\bar\lambda$. Consider \eg\ 
$\partial_\lambda^2\partial_{\bar\lambda}Z$. 
This is a three point function with two insertions of
$\int d^2z\int d^2\theta \exp\left[-(2/\al Q)(\phi+ix)\right]$
and one 
$\int d^2z\int d^2\bar\theta \exp\left[-(2/\al Q)(\phi-ix)\right]$.
The SCKG allows us to fix the locations of the three operators
on the sphere (\ie\ drop the $z$ integrals), and also drop two of the three
$\theta$ integrals. One possible gauge fixing is
\eqn\llz{\partial_\lambda^2\partial_{\bar\lambda}Z=-
\langle 0| e^{-{2\over\al Q}(\phi+ix)(z_1)}e^{-{2\over\al Q}(\phi+ix)(z_2)}
G_{-{1\over2}}^-\bar G_{-{1\over2}}^-e^{-{2\over\al Q}(\phi-ix)(z_3)}
|0\rangle.}
Using the fact that $G_{-{1\over2}}^-$, $\bar G_{-{1\over2}}^-$ commute
with $\exp(-{2\over\al Q}(\phi+ix))$, we can push them to the left, until
they annihilate the vacuum $\langle 0|$. Thus the amplitude vanishes.

Note that the vanishing of $Z_0$ relies on $N=2$ worldsheet supersymmetry,
and worldsheet conformal invariance, but not on spacetime supersymmetry. 
In fact, the non-extremal vacua we are discussing break all spacetime
supersymmetry. 
 
We see that to leading order in $1/\mu$ the free energy vanishes, 
in agreement with the energy-entropy relation \hagspec\ implied 
by the classical cigar geometry. To compute $1/\mu$ corrections, 
we have to examine string loop effects in the background \thermback. 
In the next section we will study the one loop correction $Z_1$
(see \zstrexp). 

\newsec{The leading $1/\mu$ correction to classical thermodynamics}

In the last section we saw that at very high energy density
the thermodynamics is degenerate -- the temperature is equal
to the Hagedorn one \temp\ independently of the energy density
$\mu$. It is thus of interest to compute subleading corrections
to the equation of state. As discussed in the introduction, one
expects the entropy-energy relation to take the form
\eqn\secorr{S(E)=\beta_HE+\alpha\log{E\over\Lambda}+
O\left({1\over E}\right),}
where $\Lambda$ is a dimensionful constant (a UV cutoff) which we 
will not keep track of below. 
Consider the canonical partition sum 
\eqn\zbelowh{Z(\beta)=\int_0^\infty dE\rho(E)e^{-\beta E}.}
Near the Hagedorn temperature one might expect $Z(\beta)$ to
be dominated by the contributions of high energy states;\foot{In
section 4 we will see that this assumption is valid slightly
{\it above} the Hagedorn temperature, but is {\it not} valid
slightly below it.}
if this is the case, one can replace $\rho(E)$ by \hagspec\ and find,
\eqn\evalzb{Z(\beta)\simeq \int dE E^\alpha e^{(\beta_H-\beta) E}\simeq
(\beta-\beta_H)^{-\alpha-1}.}
The free energy \cantherm\ is thus given by
\eqn\freeen{\beta\CF\simeq(\alpha+1)\log(\beta-\beta_H).}
The energy computed in the canonical ensemble\foot{Comparing to
\tempener\ we see that the canonical and microcanonical
ensembles are not equivalent. This is a familiar feature of 
systems with a Hagedorn density of states. We will return to it 
in section 4.} is
\eqn\enmean{E={\partial(\b\CF)\over\partial\beta}\simeq
{\alpha+1\over \b-\b_H};}
thus the free energy \freeen\ can be written as
\eqn\bfcorr{-\beta\CF\simeq(\alpha+1)\log E.}
Comparing to the expansion \cantherm\ -- \strnnn\ we see that the leading
term in the free energy arises from the torus (one loop) diagram
in the background \thermback, since it scales as $\mu^0$, like 
$Z_1$ in \zstrexp. In this section we will compute this term and  
determine $\alpha$. 

The torus partition sum in the background \thermback\ is in fact
divergent, since it is proportional to the infinite volume of the
cigar, associated with the region far from the tip, $\phi\to\infty$.
As is standard in other closely related contexts, we will
regulate this divergence by requiring that
\eqn\uvcutoff{\phi\leq \phi_{UV}.}
In the fivebrane theory, this can be thought of as introducing
a UV cutoff. This makes the partition sum finite, but the
bulk of the amplitude still comes from the region far
from the tip of the cigar. For the purpose of computing this
``bulk contribution'' one can replace the cigar by a long
cylinder with $\phi$ bounded on one side by the UV cutoff
\uvcutoff\ and on the other by the location of the tip
of the cigar. Combining \lindil\ and \strnnn\ we find that
\eqn\phibounds{{1\over Q}\log{\mu\over N}\le \phi\le\phi_{UV}.}  
Thus, the length of the cut-off cylinder is
\eqn\volcig{
L_\phi=\phi_{UV}-{1\over Q}\log{\mu\over N}=-{1\over Q}\log E+{\rm const}.}
Since we are only interested in the energy dependence, we
suppress in \volcig\ a large energy independent
contribution. Any contributions to the torus partition
sum from the region near the tip of the cigar can also
be lumped into this constant. Note the minus sign in front
of $\log E$ in \volcig. The length $L_\phi$ is of course
positive; the minus sign simply means that $L_\phi$
decreases as $E$ grows. 

To recapitulate, for the purpose of calculating the bulk contribution
to the torus partition sum, we can replace the background \thermback\
by
\eqn\backggg{\IR_{\phi}\times S^1\times SU(2)_N\times\IR_5.} 
The linear dilaton direction is regulated as in \phibounds.
The circumference of the $S^1$ is $\beta_H$ \temp.

The background \backggg\ is easy to analyze since it is very
similar to that describing flat space at finite temperature
(see \eg\ \refs{\mc,\brien,\att}). The bosonic fields on the
worldsheet are seven free fields, one of which
(Euclidean time) is compact, and a level $N-2$ $SU(2)$ WZW model.
The worldsheet fermions are free and decoupled from the bosons;
their partition sum, and in particular the sum over spin
structures, is the same as in the flat space analysis,
which we briefly review next.

Collecting all the contributions to the thermal torus partition sum
in the background \backggg\ we find,\foot{We follow the conventions of 
\att, which should be consulted for additional details. We also drop
the subscript $H$ on $\beta_H$, and will reinstate it later.}  
\eqn\part{\eqalign{
&Z_1={\beta V_5L_\phi\over 4}\int_F{d^2\tau\over\tau_2}
\left({1\over4\pi^2\al\tau_2}\right)^{7/2}{1\over|\eta(\tau)|^{10}}
Z_{N-2}(\tau)\times\cr 
&\sum_{n,m\in Z}\sum_{\mu,\nu=1}^4\delta_\mu U_\mu(n,m)\delta_\nu
U_\nu(n,m) \left({\vartheta}_\mu(0,\tau)\over \eta(\tau)\right)^4
\left({\vartheta}_\nu(0,\bar\tau)\over \eta(\bar\tau)\right)^4
e^{-S_\beta(n,m)}. }}
The modular integral runs over the standard fundamental domain $F$.
$Z_{N-2}$ is the partition sum of level $N-2$ $SU(2)$ WZW~\foot{We
choose the $A$ series modular invariant; the $D$ and $E$ series modular
invariants can also be studied and correspond to other vacua of LST \abks.}
(see for example \difra),
\eqn\su{
Z_{N-2}(\tau)=\sum_{m=0}^{N-2}\chi_m^{(N-2)}(q)\chi_m^{(N-2)}(\bar q)
=\sum_{m=0}^{N-2}|\chi_m^{(N-2)}(q)|^2,}
where $q=\exp(2\pi i \tau)$ and
\eqn\ch{
\chi_m^{(N-2)}(q)={q^{(m+1)^2\over 4N}\over \eta(q)^3}\sum_{n\in Z}
[1+m+2nN)]q^{n(1+m+Nn)}.}
We note for future reference that $Z_{N-2}$ is real and positive.

$\mu,\,\nu$ denote the spin structure for left and right moving 
worldsheet fermions, respectively. $\delta_\mu=(\pm,-,+,-)$ are 
signs coming from the usual GSO projections for IIA and IIB 
superstrings at zero temperature; $n,\,m$ are winding numbers of 
Euclidean time around the two non-contractible
cycles of the torus. The soliton factor $S_\beta(n,m)$ is given by
\eqn\soliton{
S_\beta(n,m)={\b^2\over 4\pi\al\tau_2}(m^2+n^2|\tau|^2-2\tau_1 mn).}
$U_\mu(n,m)$ are additional signs that are associated with finite
temperature. Their role is to implement the standard thermal boundary
conditions, that spacetime bosons (fermions) are (anti-)periodic around
the Euclidean time direction. One can show \att\ that this requirement 
together with modular invariance leads to:
\eqn\us{
\eqalign{
&U_1(n,m)={1\over 2}\left (-1+(-1)^n+(-1)^m+(-1)^{n+m}\right)\cr
&U_2(n,m)={1\over 2}\left (1-(-1)^n+(-1)^m+(-1)^{n+m}\right)\cr
&U_3(n,m)={1\over 2}\left (1+(-1)^n+(-1)^m-(-1)^{n+m}\right)\cr
&U_4(n,m)={1\over 2}\left (1+(-1)^n-(-1)^m+(-1)^{n+m}\right).}}
The terms with $\mu=1$ in \part\ vanish because of the presence 
of fermionic zero modes for the $(+,\,+)$ spin structure, 
or equivalently since $\vartheta_1(0,\tau)=0$.

The torus partition sum \part\ can be rewritten in
a way that makes it manifest that the coefficient of
$\beta V_5 L_\phi/4$ is positive,
\eqn\neg{\eqalign{
&Z_1={\beta V_5 L_\phi\over 4}\int_F{d^2\tau\over\tau_2}\left
({1\over
4\pi^2\al\tau_2}\right)^{7/2}{1\over|\eta(\tau)|^{18}}
Z_{N-2}(\tau)\times\cr 
&\sum_{n,m\in Z}\left|\sum_{\mu=2}^4 U_\mu(n,m)\delta_\mu
{\vartheta}_\mu^4(0,\tau)\right|^2
e^{-S_\beta(n,m)}.}}
It is not difficult to check that the integral \neg\ is
convergent at $\tau_2\to\infty$, the only region where
a divergence could occur. 

To exhibit the interpretation of \neg\ as a sum over the
free energies of physical string modes one can proceed as
follows \refs{\polch,\mc,\brien}. Using
the modular invariance of the integrand and the
covariance of $(n,m)$, one can
extend the integral from the fundamental domain to the strip
\eqn\st{
S:\quad-{1\over 2}\leq\tau\leq{1\over 2};\quad \tau_2\geq 0,}
while restricting to configurations with $n=0$ in \neg.
This leads to 
\eqn\nega
{\eqalign{
&Z_1={\beta V_5L_\phi\over 4}\int_{S}{d^2\tau\over\tau_2}\left
({1\over
4\pi^2\al\tau_2}\right)^{7/2}{1\over|\eta(\tau)|^{18}}
Z_{N-2}(\tau)\times \cr
&\sum_{m=-\infty}^\infty\left|\sum_{\mu=2}^4 U_\mu(0,m)\delta_\mu
{\vartheta}_\mu^4(0,\tau)\right|^2
e^{-S_\beta(0,m)}.}}
The integral over $\tau_1$ projects on physical states (\ie\ those
with $L_0=\bar L_0$), while $\tau_2$ plays the role of a Schwinger
parameter. Because of the Jacobi identity
$\vartheta_2^4(0,\tau)-\vartheta_3^4(0,\tau)+\vartheta_4^4(0,\tau)=0$,
and the fact that $U_2(0,m)=(-)^m$, $U_3(0,m)=U_4(0,m)=1$, 
the sum over $m$ in \nega\ can be restricted to odd integers. 
It is not difficult to check in this representation too that the 
integral over $\tau_2$ is convergent.

We are now ready to determine the parameter $\alpha$ in \secorr,
\bfcorr. Using the relation \cantherm\ between the free energy
$\CF$ and the string partition sum, as well as \bfcorr, we see 
that $Z_1$ should be proportional to $\log E$. This is indeed
the case in \nega\ since the length $L_\phi$ goes like
$-\log E$ (see \volcig). Combining these relations we find
that
\eqn\formalpha{
\eqalign{
&\alpha+1=-{\beta V_5\over 4Q}\int_{S}{d^2\tau\over\tau_2}\left
({1\over
4\pi^2\al\tau_2}\right)^{7/2}{1\over|\eta(\tau)|^{18}}
Z_{N-2}(\tau)\times \cr
&\sum_{m=-\infty}^\infty\left|\sum_{\mu=2}^4 U_\mu(0,m)\delta_\mu
{\vartheta}_\mu^4(0,\tau)\right|^2
e^{-S_\beta(0,m)}.}}
We see that $\alpha+1$ is negative, as stated above.\foot{Of course,
since the r.h.s. of \formalpha\ is proportional to $V_5$ which
is assumed to be very large, we can neglect the $+1$ on the left hand
side.}
Physically, it is clear that it is counting the free energy of the
perturbative string modes which live in the vicinity of the
black brane. An interesting point which was mentioned in 
\refs{\berk,\har}
is that $\alpha$ is an extensive quantity -- it is proportional
to the volume of the fivebrane $V_5$, in contrast, say, to the
one particle free energy in critical string theory, where the 
analogous quantity is of order one.

The integral \formalpha\ appears in general to be rather formidable 
and we do not know whether it can be performed exactly. In the remainder 
of this section we will compute it in the limit $N\to\infty$, where
the computation simplifies.


For large $N$ the partition sum corresponding 
to the three-sphere, $Z_{N-2}(\tau)$, simplifies significantly.
Indeed, for $N\gg 1$ \su\ can be approximated as 
\eqn\Z{Z_{N-2}(\tau)={1\over |\eta(q)|^6}\sum_{p=0}^\infty
|q|^{(p+1)^2\over 2N}(p+1)^2.}

Returning to the evaluation of $\alpha$, \formalpha, we have
\eqn\parto{
\eqalign{
&\alpha+1=-{\beta V_5\over 4Q}
\left({1\over 4\pi^2\al}\right)^{7/2}
\int_S{d^2\tau\over\tau_2^{9/2}}\left|{1\over \eta(\tau)}\right|^{24}\times\cr
&\sum_{m\in 2Z+1}\sum_{p=0}^{\infty}
e^{-{(p+1)^2\tau_2\over 2N}}(p+1)^2 e^{-{\b^2m^2\over 4\pi\al\tau_2}}
\left|\vartheta_2^4+\vartheta_3^4-\vartheta_4^4\right|^2(0,\tau).}}
At this point it is useful to recall that the inverse temperature
$\beta$ in \parto\ is in fact the Hagedorn temperature of LST, \temp.
In the large $N$ limit, $\beta_H\sim\sqrt N$ becomes large (or,
equivalently, the Hagedorn temperature is small in string units) and the 
exponential term in \parto\ suppresses the amplitude, unless $\tau_2$ 
is large as well (of order $N$). 
Therefore, the $\tau$ integral in \parto\ is
dominated by the large $\tau_2$ region, which corresponds to the
free energy of the supergravity modes. To compute the integral
we recall the asymptotic forms of the $\vartheta$ and $\eta$ functions
at large $\tau_2$ (see \eg\ \polchin)
\eqn\thet{\eqalign{
&\vartheta_2(0,\,\tau)=\sum_{n=-\infty}^\infty q^{{1\over2}(n- {1\over
2})^2}=2q^{1\over 8}(1+q+\dots)\cr 
&\vartheta_3(0,\,\tau)=\sum_{n=-\infty}^\infty q^{{1\over2}n^2}=
1+2q^{1\over 2}+\dots\cr
&\vartheta_4(0,\,\tau)=\sum_{n=-\infty}^\infty(-1)^n q^{{1\over2}n^2}
=1-2q^{1\over 2}+\dots\cr
&\eta(\tau)=q^{1\over 24}\prod_{n=1}^{\infty}(1-q^n)=
q^{1\over 24}+\dots\,.
}}
Plugging in \parto\ and using the definition of the modified Bessel function
\eqn\modbes{
K_\nu(z)={1\over 2}\left(2\over z\right)^\nu\int_0^\infty t^{\nu-1}e^{-{z^2\over 4t}-t}dt,
}
we find
\eqn\free{\eqalign{
\alpha+1=&-{8V_5\over\pi^6(N\al)^{5/2}}\sum_{k,p=0}^\infty
\left({2\pi(2k+1)^2\over (p+1)^2}\right)^{-7/4}(p+1)^2\times\cr 
&K_{-{7\over 2}}(\sqrt{2\pi}(p+1)(2k+1))\simeq
-{4.08\cdot 10^{-4}V_5(N\al)^{-5/2}}\equiv -a_1 V_5.
}}
 
Note that, as expected, $a_1$ is positive. Of course, as is clear 
from \parto, we can write $\alpha+1$ as $-a_1 V_5$ with $a_1$ 
a positive constant for all $N$, but in general $a_1$ receives contributions
from massive string modes and is thus given by a complicated modular integral.
The large $N$ behavior of $a_1$ is simpler and is given by \free.

The fact that $\alpha$ goes like $N^{-5/2}$ for large $N$ was found in a
different way in \berk, by analyzing the deformation of the classical
solution \metro\ at  one string loop. Our analysis determines the 
coefficient of $N^{-5/2}$, and in particular its sign which,
as mentioned above, is important for the thermodynamics.

In the discussion above, the fivebrane was assumed
to be effectively non-compact. It is interesting to study the 
thermodynamics of fivebranes wrapped around compact manifolds, 
and in particular the dependence of $\alpha$ on the size and 
shape of the manifold. As an example of the sort of dependence 
one can expect, consider compactifying the fivebrane on $(S^1)^5$ 
where all five circles have the same radius $R$. It is sufficient
to consider the case $R\ge\sqrt{\alpha'}$ since smaller radii give rise
to the same physics due to T-duality. 

As is standard in string theory, the effect of this is to replace
the contribution of the non-compact zero modes on $R^5$ by the 
momentum and winding sum on $(S^1)^5$:
\eqn\partfree{
{V_5\over (4\pi^2\al\tau_2)^{5/2}}\longrightarrow
\left(\sum_{l,p\in Z}
q^{{\al\over 4}\left({l\over R}+{pR\over\al}\right)^2}
{\bar q}^{{\al\over 4}\left({l\over R}-{pR\over\al}\right)^2}\right)^5.
}
Consider for simplicity the limit $N\to\infty$ discussed above. 
As mentioned after eq. \parto, since the Hagedorn temperature
is very low, the modular integral is dominated in
this case by $\tau_2\sim N$. If the radius $R$ is much larger than
$\sqrt{N\alpha'}$, the sum over momenta on the r.h.s. of \partfree\ 
can be approximated by an integral and gives the same contribution
as in the non-compact case (namely the l.h.s. of \partfree). 
For $R\sim \sqrt{N\alpha'}$ one has to include a few low lying
momentum modes -- this is a transition region. For
$\sqrt{\alpha'}<R\ll\sqrt{N\alpha'}$ one can neglect all contributions of
momentum (and winding) modes just like one is neglecting the contributions of
oscillator states. Thus, we get in this case
\eqn\newal{\eqalign{
&\alpha+1=-{\b\over 2Q}\left({1\over 4\pi^2\al}\right)
\int_0^\infty {d\tau_2\over\tau_2^2}\cdot 1024\sum_{k,p=0}^\infty
e^{-{\b^2 (2k+1)^2\over 4\pi\al\tau_2}-
{(p+1)^2\tau_2\over 2N}}=\cr
&-{256\over\pi}\sum_{k,p=0}^\infty
\left({2\pi(2k+1)^2\over (p+1)^2}\right)^{-1/2}(p+1)^2
K_{-1}(\sqrt{2\pi}(p+1)(2k+1))\simeq
-3.693.
}}
Interestingly, we find that for small fivebranes $\alpha$ is 
independent of the number of fivebranes $N$ in the 
$N\rightarrow\infty$ limit. Note also that in this case
it is important to keep the $+1$ on the l.h.s. of \newal,
since $\alpha$ is of order one.

To summarize, the power $\alpha$ that appears in the high energy
density of states \enentr\ exhibits an interesting dependence
on the size of the spatial manifold that the fivebranes are
wrapping. For manifolds of size much larger than the characteristic scale
of LST, $\sqrt{N\alpha'}$, $\alpha$ is proportional to the volume of
the manifold, while for sizes much smaller than this charateristic
scale, it saturates at a finite value, which is independent of $N$ 
(for large $N$),
\newal. If the density of states \enentr\ is due to strings confined
to the fivebranes, then these strings belong to a new universality
class, with typical configurations not exceeding the size 
$\sqrt{N\alpha'}$. It would be interesting to understand this
universality class better (see also \berk).

\newsec{Comments on the near-Hagedorn thermodynamics of LST}

The main result of our discussion so far is that the
thermodynamics corresponding to non-extremal fivebranes
is unstable. The temperature-energy relation has the form
\tempener, with $\alpha$ given by \parto\ or for large $N$
by \free, \newal. Since it is negative, the temperature is 
above the Hagedorn temperature, and the specific heat is 
negative. This raises two immediate questions:
\item{(1)} What is the thermodynamics for temperatures
slightly below the Hagedorn temperature? 
\item{(2)} What is the nature of the instability above
the Hagedorn temperature?

\noindent
The purpose of this section is to discuss these issues.  
Consider first the behavior well below the Hagedorn temperature,
$\beta\gg\beta_H$.
In this regime, the thermodynamics is expected to reduce to that  
corresponding to the extreme IR limit of LST, which is the
$(2,0)$ six dimensional SCFT for type IIA LST, or six dimensional
$(1,1)$ SYM for IIB. From the point of view of the holographic
description, this regime corresponds to the strong coupling
region of the near-horizon geometry of the fivebranes \abks, 
and thus should not be visible in the perturbative theory on 
the cigar \metro.

What happens as the temperature approaches $T_H$ from below?  
One might expect that due to the Hagedorn growth in the density 
of states \enentr, the high energy part of the  spectrum dominates 
as $\beta\to\beta_H$, and the partition sum is well approximated by 
\evalzb. What actually happens depends on the value of $\alpha$, 
as we discuss next. 

Consider first the case of large $V_5$ ($R\gg\sqrt{N\alpha'}$ in the
discussion at the end of section 3). In this case, $|\alpha|$ is large,
and the contribution to the partition sum of the high energy part of
the spectrum, \evalzb,
goes rapidly to zero as $\beta\to\beta_H$. The integral over $E$ is
dominated by  states with moderate energies, whose contribution  to the
partition  sum is analytic at $\beta_H$. It is clear that the mean
energy remains finite as we approach the Hagedorn temperature from below,
and that thermodynamic fluctuations are suppressed (by a factor of the
volume $V_5$). Since the Hagedorn temperature is reached at
a finite energy, it corresponds to a phase transition.

As $V_5$ decreases, $\alpha$ decreases as well, until it reaches the
value \newal. The fluctuations in energy in the canonical ensemble
increase with decreasing $\alpha$. To see that, 
consider the case $R\ll\sqrt{N\alpha'}$ in the discussion at the end 
of section 3. Since $-5<\alpha<-4$ in that case, the expectation
values $\langle E^n\rangle$ with $n\ge 4$ 
in the canonical ensemble diverge as
\eqn\mmeeaa{\langle E^n\rangle \sim (\beta-\beta_H)^{-\alpha-n-1}.}
In such situations, one is instructed to pass to the
microcanonical ensemble, in which the energy is fixed and the
temperature is defined by \tempener. The perturbative evaluation
of $\beta$ in \tempener\ gives a temperature {\it above}
the Hagedorn temperature. This of course does not imply that LST
cannot be defined at temperatures below $T_H$; instead, it means
that to study the theory at such temperatures one must compute 
$S(E)$ to all orders in $1/E$, include non-perturbative
corrections, and solve the equation
$$\beta={\partial S(E)\over \partial E}$$
to find the energy $E$ corresponding to a particular $\beta>\beta_H$.
From the form of the leading terms in $S(E)$ it is clear that the
solution of this equation will correspond to finite $E$. We are led
again to the conclusion that 
the Hagedorn temperature is reached at a finite
energy and thus is associated with a phase transition.
  
Since the study of the non-extremal fivebrane geometry in the previous
sections is perturbative in $1/E$, it is not useful for studying the regime
$\beta>\beta_H$. Nevertheless, it seems clear that the  specific heat is
positive there (this is certainly the case for the infrared theory on the
fivebranes). Furthermore, since the energy -- temperature relation is such that
the Hagedorn temperature is reached at a finite energy, we are led to the
second question raised in the beginning of this section: what is the nature of
the high temperature phase of LST?

The perturbative analysis of the near-extremal fivebrane, which
is valid for $\beta$ slightly below $\beta_H$, predicts that the
thermodynamics is unstable. Usually, in such situations the
instability is associated with a negative mode in the Euclidean
path integral (a tachyon). Examples include the instability of flat space
at finite temperature in Einstein gravity \gpy, and the thermal tachyon
that appears above the Hagedorn transition in critical string theory. The
one loop instability found in section 3 leads one to believe that a similar
negative mode should appear in LST above the Hagedorn temperature.

At first sight this statement appears surprising. For large but finite $N$,
far from the tip of the cigar \metro, the near-horizon geometry \thermback\
is essentially the same as in critical string theory at the temperature
$T_H$ \temp, which goes to zero in the limit $N\to\infty$. There are clearly
no tachyons in critical string theory at low temperature; thus we conclude
that any unstable modes of the thermal background \thermback\ must be localized
near the tip of the cigar, \ie\ near $\sigma=0$ in the coordinates \metro.
This is natural, since one expects a phase transition to change
the structure of the horizon of the black brane; the asymptotic 
behavior far from the tip of the cigar (a cylinder with
circumference $\beta$) should remain unchanged.

States localized near the tip of the cigar in LST were studied
in \GK. A convenient way to study them is to construct observables
which correspond to vertex operators whose wave-functions are
non-normalizable at $\phi\to\infty$ (which can be thought of as
off-shell observables in LST) and compute their correlation functions.
Normalizable states on the cigar, which are created by these observables
acting on the vacuum, give rise to poles in these correlation functions.
The masses of these states can be read off the locations of the poles.

In our case, it turns out that the observable that creates the light
state when acting on the vacuum is the ``fermionic string tachyon,''
whose vertex operator in the $(-1,-1)$ picture is
\eqn\fertach{\CT_m(\vec p)=e^{-\varphi-\bar\varphi}
V_{j;m,m}e^{i\vec p\cdot\vec x}.}
The notation here is the following (see \GK\ for further details). 
$\varphi$, $\bar\varphi$  are the bosonized superconformal
ghosts. $\vec p$ is
the spatial momentum along the fivebrane. $V_{j;m,m}$
is an observable in the cigar CFT. It belongs to a large class
of primaries in the cigar SCFT corresponding to momentum 
and winding modes $V_{j;m,\bar m}$, where $(m,\bar m)$ are related
to the momentum and winding numbers around the cigar, $n, w\in Z$, 
via
\eqn\momwin{m={1\over2}(n+wN);\;\;\bar m=-{1\over2}(n-wN).}
The worldsheet scaling dimensions of these observables are
\eqn\scdimv{\Delta_{j;m,\bar m}={m^2-j(j+1)\over N};\;\;
\bar\Delta_{j;m,\bar m}={\bar m^2-j(j+1)\over N}.}
In particular, we see that the observable \fertach\ corresponds
to a pure winding mode around Euclidean time, with winding number
$w=2m/N$.  

The mass shell condition for the vertex operator \fertach\ is
\eqn\mmaass{{\alpha'\over4}|\vec p|^2+{m^2-j(j+1)\over N}={1\over2}.}
This equation can be thought of as determining $j$ as a function of
$|\vec p|$ and $m$. The operators $\CT_m(\vec p)$ thus correspond to 
off-shell observables in LST \GK.

Not all observables \fertach\ are physical. The GSO projection implicit
in the partition sum \part\ projects out those with even winding number
$w$, leaving behind those with $w\in 2Z+1$. This is analogous to the
situation in flat space where the finite temperature GSO projection
projects out tachyons with even winding number. This analogy suggests
that the operator \fertach\ with winding number one creates from the
vacuum the thermal tachyon for $\beta<\beta_H$. We will next show that
this is indeed the case.

As explained in \GK, the two point function of the operator \fertach\
has a simple pole whenever $m$ and $j$ belong to a principal discrete
series representation\foot{There are bounds on $j$ which are discussed
in \GK; they are satisfied here and will not be reviewed.} of $SL(2)$,
\ie\ for
\eqn\mmmjjj{m=j+l;\;\;l=1,2,3,\cdots\,.} 
The lowest mass state corresponds to
$l=1$; plugging into \mmaass\ we see that the corresponding 
mass-shell condition is  
\eqn\massbound{{\alpha'\over4}|\vec p|^2+{w-1\over2}=0.} 
Thus, the mass is
\eqn\almass{{\alpha'\over4}M_w^2={w-1\over2}.} 
The mode with winding number zero would have corresponded to a tachyon had 
it existed, but it is projected out by GSO. The winding number
one ($w=1$) mode is massless; the higher (odd) winding number modes are
massive.  It is not difficult to show that the masses of all states 
obtained by repeating the above procedure for other observables
are strictly positive.

Since the vacuum we are studying is non-supersymmetric, it is
reasonable to expect that the masses \almass\ receive quantum 
corrections. In particular, it is likely that the classically
massless state with $w=1$ is lifted at one loop
(see \zstrexp, \strnnn):
\eqn\masswone{M_1^2=C e^{2\Phi_0} +O(e^{4\Phi_0})= 
{CN\over\mu}+O({1\over\mu^2}).}
We have not computed these corrections, but would like to argue
that $C<0$, so that string loop effects drive the massless
state tachyonic. This would lead to a consistent picture of
the high temperature phase of LST. The perturbative thermodynamics
is marginally stable classically \hagspec, and is destabilized at
one loop \enentr. Correspondingly, the classical Euclidean
theory contains a zero mode winding once around Euclidean time, and
one loop effects turn it into a negative mode.

Needless to say,
it would be interesting to verify the conjecture that $C$ is
negative by an explicit one loop calculation. Assuming that this
is indeed the case, we arrive at the picture described earlier
in the paper: the fivebranes reach the Hagedorn temperature at a
finite energy density. At that point the mode with winding number
one described above turns tachyonic and condenses. The system thus
undergoes a phase transition.

The precise temperature at which this condensation occurs depends
on the behavior of the tachyon potential for
$\beta\simeq\beta_H$. The quadratic term in the potential
was argued above to change sign at $\beta_H$.
The physics of the phase transition depends on the sign of the quartic
term. It would be interesting to compute this term directly.
It would also be nice to describe the endpoint of tachyon condensation.
Since our description of this mode is rather indirect (as a pole in a
correlation function of the observables \fertach), and the condensation
involves understanding string loop effects, this remains an
interesting challenge.

\bigskip
\noindent{\bf Acknowledgements:}
We thank O. Aharony for useful discussions and comments on the manuscript. 
This work was supported in part by DOE grant \#DE-FG02-90ER40560. 

\listrefs
\bye